# ANNs-SaDE: A Machine-Learning-Based Design Automation Framework for Microwave Branch-Line Couplers

Tianqi Chen[1]  Wei Huang[1]  Qiang Wu[1]  Li Yang[2]  Roberto Gómez-García[2]  Xi Zhu[1]
[1]School of Electrical and Data Engineering, University of Technology Sydney, Ultimo, Australia
[2]Dept of Signal Theory and Communications, University of Alcalá, Alcalá de Henares, Madrid, Spain

*Abstract*—The traditional method for designing branch-line couplers involves a trial-and-error optimization process that requires multiple design iterations through electromagnetic (EM) simulations. Thus, it is extremely time consuming and labor intensive. In this paper, a novel machine-learning-based framework is proposed to tackle this issue. It integrates artificial neural networks with a self-adaptive differential evolution algorithm (ANNs-SaDE). This framework enables the self-adaptive design of various types of microwave branch-line couplers by precisely optimizing essential electrical properties, such as coupling factor, isolation, and phase difference between output ports. The effectiveness of the ANNs-SaDE framework is demonstrated by the designs of folded single-stage branch-line couplers and multi-stage wideband branch-line couplers.

## I. INTRODUCTION

Quadrature couplers are essential RF passive components in various wireless systems, including balanced amplifiers (BAs) [1], [2], Doherty-like power amplifiers (PAs) [3], [4], and I/Q generators [5], [6]. These couplers can be designed either in coupled-line or branch-line structures, depending on the required design specifications. In planar circuits, branch-line structures are often preferred over coupled-line arrangements, because they do not require strong coupling levels between lines to attain 3-dB coupling factors, which is challenging to achieve with a single metal layer. On the other hand, the traditional method for designing passive components involves device modeling using equivalent circuits and then iteratively optimizing them by means of electromagnetic (EM) simulation tools. This process can be quite time-consuming, often taking from several hours to even days to achieve results that satisfy the prefixed design specifications. The labor-intensive nature of this approach underscores the urgent need for design automation driven by artificial intelligence (AI) in both industry and academia.

Recently, the design automation for quadrature couplers has gained significant attention. Despite some advancements, the research of AI-based design automation for quadrature couplers is still in a relatively infancy stage. Until now, only a few examples have been reported in the technical literature. In [7], millimeter-wave directional couplers are designed by an ANNs-based surrogate model with backpropagation optimization algorithm. Optimizations of substrate-integrated waveguide (SIW) couplers with varying power-division ratio are discussed through a Q-Network [8]. Different couplers with minimal random data are optimized by two-stage inverse models [9]. A co-kriging surrogate model is presented to support coupler design automation [10]. However, all these presented algorithms in [7]–[10] exclude the optimization of the phase difference between the output ports of the coupler. To address this issue, a novel optimization framework is proposed by combining the artificial neural networks with a self-adaptive differential evolution algorithm (ANNs-SaDE). The developed framework can determine the physical parameters of EM structures to satisfy critical electrical properties without requiring additional EM simulations. The rest of this paper is organised in the following way. In Section II, the proposed methodology is discussed in detail. Afterwards, two design examples are graphically presented in Section III. Finally, the conclusion is drawn in Section IV.

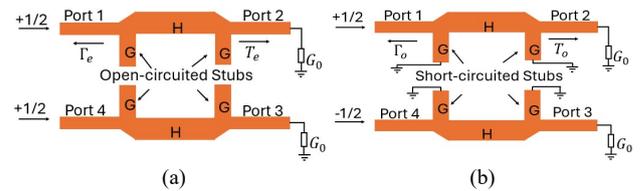

Fig.1. Classical branch-line coupler. (a) Even mode. (b) Odd mode.

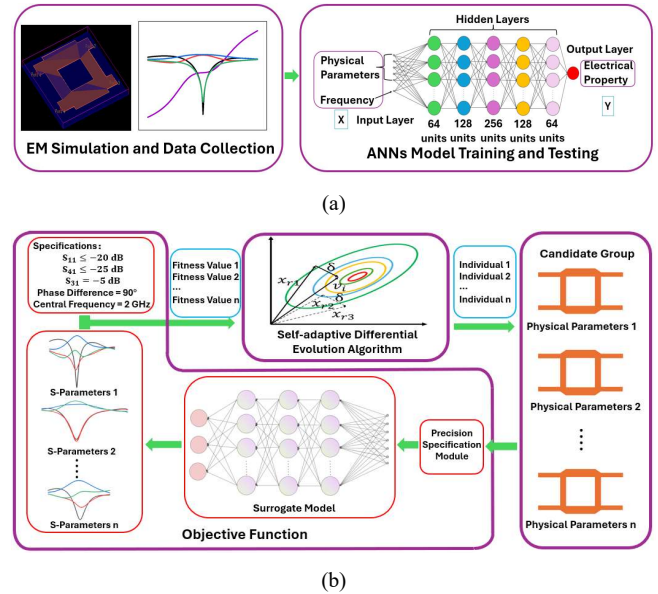

Fig. 2. Proposed ANNs-SaDE framework.

## II. PROPOSED METHODOLOGY

### A. Theoretical Analysis of a Classical Branch-line Coupler

The even-odd mode analysis approach is utilized to obtain the electrical properties of branch-line couplers in Fig. 1. The *S*-parameters of the branch-line coupler are as follows:

$$S_{11} = \tfrac{1}{2}\Gamma_e + \tfrac{1}{2}\Gamma_o \qquad (1)$$

$$S_{21} = \tfrac{1}{2}T_e + \tfrac{1}{2}T_o \qquad (2)$$

$$S_{31} = \tfrac{1}{2}T_e - \tfrac{1}{2}T_o \qquad (3)$$

$$S_{41} = \tfrac{1}{2}\Gamma_e - \tfrac{1}{2}\Gamma_o \qquad (4)$$

where the $\Gamma_e$, $T_e$, $\Gamma_o$, and $T_o$ are the reflection and transmission coefficients of the even-mode circuit and odd-mode circuit, respectively. The transmission matrix of the even-mode and odd-mode circuits can be calculated by multiplying the ABCD matrices of each component as follows:

$$\begin{bmatrix} A & B \\ C & D \end{bmatrix}_e = \begin{bmatrix} -\frac{G}{H} & j\frac{1}{H} \\ jH - j\frac{G^2}{H} & -\frac{G}{H} \end{bmatrix} \qquad (5)$$

$$\begin{bmatrix} A & B \\ C & D \end{bmatrix}_o = \begin{bmatrix} \frac{G}{H} & j\frac{1}{H} \\ jH - j\frac{G^2}{H} & \frac{G}{H} \end{bmatrix} \qquad (6)$$

$$\Gamma_e = \frac{j(\frac{G^2}{H} + \frac{1}{H} - H)}{-\frac{2G}{H} + j(H + \frac{1}{H} - \frac{G^2}{H})} \qquad (7)$$

$$T_e = \frac{2}{-\frac{2G}{H} + j(H + \frac{1}{H} - \frac{G^2}{H})} \qquad (8)$$

$$\Gamma_o = \frac{j(\frac{G^2}{H} + \frac{1}{H} - H)}{\frac{2G}{H} + j(H + \frac{1}{H} - \frac{G^2}{H})} \qquad (9)$$

$$T_o = \frac{2}{\frac{2G}{H} + j(H + \frac{1}{H} - \frac{G^2}{H})}. \qquad (10)$$

where $H$ and $G$ denote the normalized admittance of the through arm and the coupling arm, respectively. Moreover, the design conditions of perfect matching for port1 and perfect isolation between port1 and port4 are imposed, which means $S_{11} = 0$ and $S_{41} = 0$. Based on (1)–(4) and (7)–(10), these conditions result in the following relationships:

$$G^2 = H^2 - 1 \qquad (11)$$

$$\Gamma_o = \Gamma_e = 0 \qquad (12)$$

$$T_e = \frac{H}{-G + j} \qquad (13)$$

$$T_o = \frac{H}{G + j} \qquad (14)$$

With the definition of coupling factor, it can be determined as follows:

$$C(dB) = 20\lg\left|\frac{1}{S_{31}}\right| = 20\lg\frac{G^2 + 1}{GH}. \qquad (15)$$

According to (11) and (15), the physical parameters of classical branch-line couplers with different coupling factors can be calculated [11]. Although the presented closed-form equations can be used to guide branch-line coupler design using the conventional EM structure, such as the one shown in Fig. 1, there are some limitations in this approach. In practice, the physical dimensions of a branch line coupler need to be changed when different design specifications are required. One of the typical user cases is to fold the branch-line coupler for a miniaturization design. In this case, a trail-and-error approach is needed for performance optimization. Furthermore, to expand the bandwidth, multi-stage branch-line couplers are used. The theoretical analysis of multi-stage branch-line couplers has become quite complex. Moreover, to maximize the bandwidth, circuits often need to operate in a slightly-mismatched state. This significantly increases the uncertainty in circuit design, while the proposed model provides a feasible solution to this uncertainty.

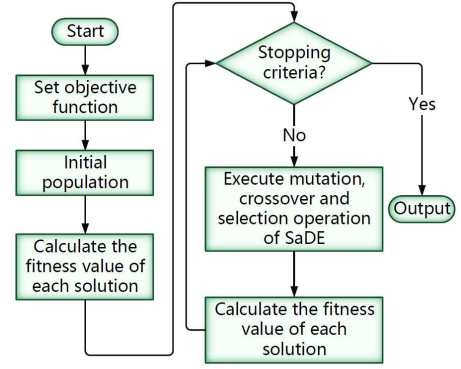

Fig. 3. Flowchart used in the discovery phase.

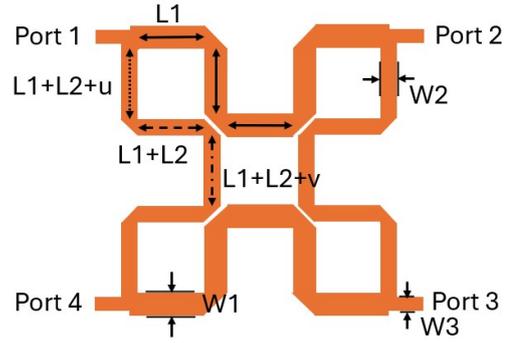

Fig. 4. Physical parameters (including the range) of the exemplified folded branch-line coupler. Note: all units are in mm. W1 is from 1.5 to 3, L1 is from 5 to 12, W2 is from 0.3 to 2, L2 is from 0.1 to 2, W3 is from 1.2 to 1.8, u = 0.1 mm, and v = 1 mm.

*B. Framework of ANNs-SaDE*

The proposed ANNs-SaDE framework consists of two different phases: the training phase and the discovery phase, as illustrated in the Fig. 2. More details are given below:

1. Training Phase: In this phase, artificial neural networks (ANNs) serving as surrogate models are trained by sampling data from the design space. The specific structure of the ANN used in this work is detailed in Fig. 2(a).

2. Discovery Phase: During this phase, the trained surrogate model is incorporated into the objective function, and the self-adaptive differential evolution (SaDE) algorithm [12] is employed to search the optimal solution within the specified range of physical parameters. The process for the discovery phase is illustrated in Fig. 3.

*C. Objective Function Design*

The objective function is formulated to determine the optimal physical parameters for branch-line couplers. The objective function $f(x)$ can be defined as

$$f(x) = \alpha L_{CF}(x) + \beta L_{PD}(x) + \gamma L_{IS}(x) + \lambda L_{RC}(x). \qquad (16)$$

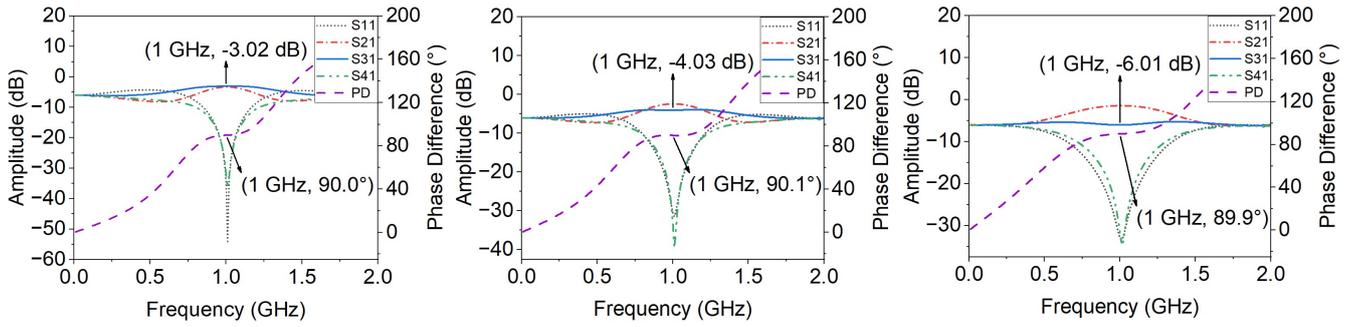

Fig. 5. Frequency responses of 1-GHz folded branch-line coupler in Fig. 4 with different coupling factor.

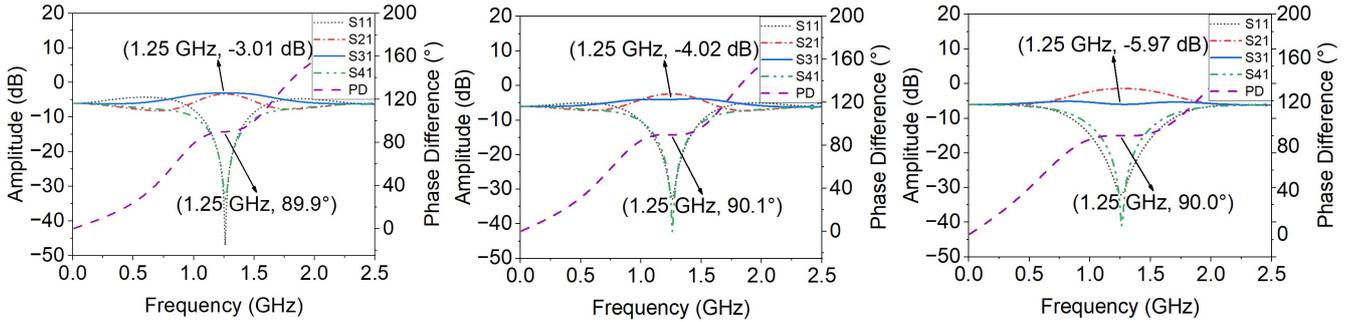

Fig. 6. Frequency responses of 1.25-GHz folded branch-line coupler in Fig. 4 with different coupling factor.

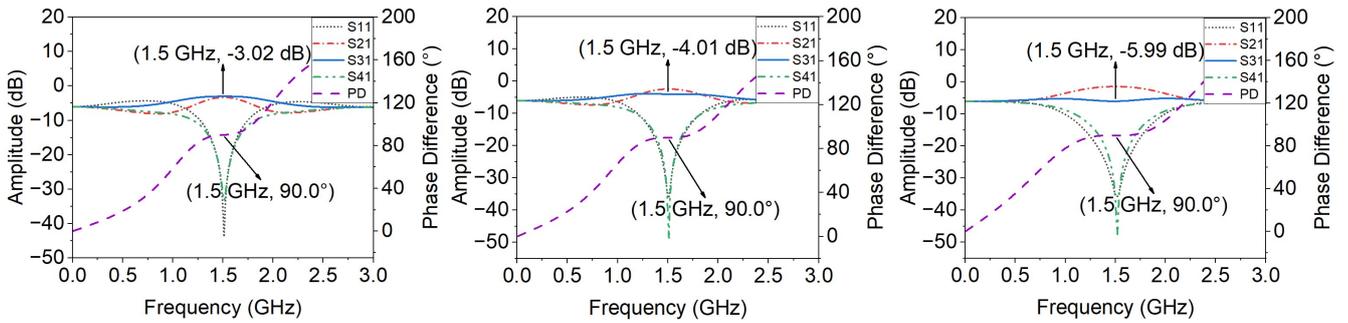

Fig. 7. Frequency responses of 1.5-GHz folded branch-line coupler in Fig. 4 with different coupling factor.

The terms $L_{CF}$, $L_{PD}$, $L_{IS}$, and $L_{RC}$ refer to the difference between individual in population and specifications about the coupling factor, phase difference between the output ports, isolation, and reflection coefficients at the input port at a specific frequency with weight coefficients $\alpha$, $\beta$, $\gamma$, and $\lambda$, respectively. The global optimization algorithm aims to identify $x^*$, which meets the following defined requirements:

$$x^* = arg \min_{X \in \Omega} f(x). \quad (17)$$

## III. DESIGN EXAMPLES

### A. Design of Folded Branch-line Couplers

As the conventional branch-line coupler occupies a large area, miniaturization techniques must be developed to address this issue. The inward bending technique is a typical technique in design miniaturization. However, as previously mentioned, the miniaturized branch-line coupler cannot be directly implemented using the classical closed-form equations. Therefore, in this sub-section, a case study for design automation of folded branch-line couplers will be presented. The structure of the folded coupler is illustrated in Fig. 4. The model aims to design couplers with a center frequency from 1 to 1.5 GHz and a coupling factor from 3 to 6 dB. The selected substrate is Rogers RT/Duroid 5880, with a relative dielectric constant of 2.2, a loss tangent of 0.0009, and a thickness of 0.508 mm. The model is trained for 500 epochs on a dataset of 500 samples. Testing results on a dataset of 100 samples are summarized in Table I. Recommended physical parameters are summarized in Table II, and their *S*-parameters were validated using Keysight ADS EM solver, as shown in Figs. 5 to 7.

TABLE I
TRAINING RESULTS OF THE SURROGATE MODEL FOR FOLDED BRANCH-LINE COUPLER

| Electrical Property | MAE in Test Dataset |
|---|---|
| $\|S_{11}\|$ (dB) | 0.093 |
| $\|S_{31}\|$ (dB) | 0.030 |
| $\|S_{41}\|$ (dB) | 0.062 |
| Phase $S_{21}$ (°) | 0.481 |
| Phase $S_{31}$ (°) | 0.403 |

TABLE II
RECOMMENDED PHYSICAL PARAMETERS OF THE FOLDED BRANCH-LINE COUPLER (ALL IN MM)

| Physical Parameters | W1 | L1 | W2 | L2 | W3 |
|---|---|---|---|---|---|
| 3 dB @ 1 GHz | 2.7 | 9.1 | 1.7 | 0.8 | 1.7 |
| 4 dB @ 1 GHz | 2.5 | 9.2 | 1.3 | 1.1 | 1.8 |
| 6 dB @1 GHz | 1.9 | 9.9 | 0.6 | 1.1 | 1.2 |
| 3 dB @1.25 GHz | 2.7 | 7 | 1.7 | 0.7 | 1.8 |
| 4 dB @1.25 GHz | 2.5 | 7.2 | 1.3 | 1 | 1.4 |
| 6 dB @ 1.25 GHz | 2.1 | 7.8 | 0.7 | 0.6 | 1.2 |
| 3 dB @ 1.5 GHz | 2.7 | 5.7 | 1.7 | 0.6 | 1.4 |
| 4 dB @ 1.5 GHz | 2.5 | 5.9 | 1.3 | 0.5 | 1.8 |
| 6 dB @ 1.5 GHz | 2.1 | 6.1 | 0.7 | 1 | 1.7 |

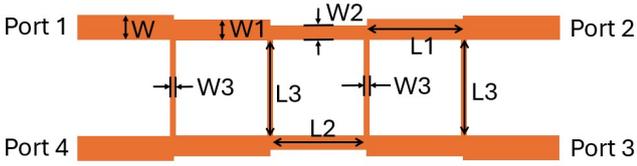

Fig. 8. Physical parameters (including the range) of the exemplified wideband cascaded coupler. Note: all units are in mm. W1 is from 3.5 to 5, L1 is from 20 to 30, W2 is from 3 to 4.5, L2 is from 20 to 30, W3 is from 0.1 to 1, L3 is from 20 to 30, and W is from 4.5 to 5.5.

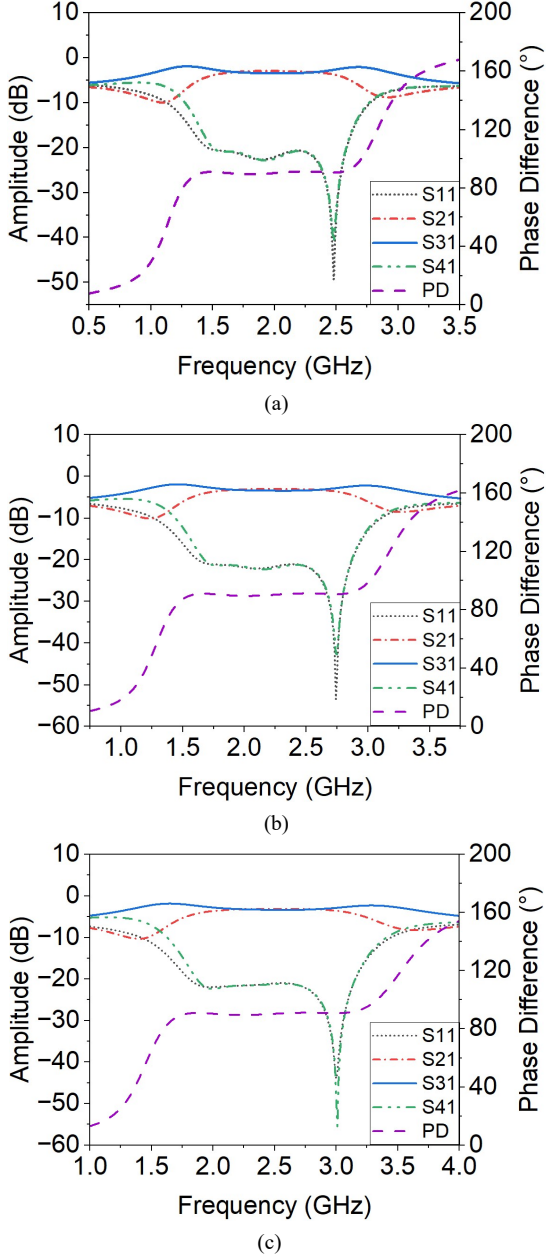

(a)

(b)

(c)

Fig. 9. Frequency responses of the wideband three-stage branch-line coupler in Fig. 8 with different center frequencies.

### B. Design of Wideband Branch-Line Couplers

In addition to varying coupling factors, many applications require couplers to provide a wide bandwidth. The most effective method for achieving this is to cascade multiple single-stage branch-line couplers. However, when cascading is applied, the closed-form expressions for design and optimization of wideband couplers become cumbersome. This study designs wideband branch-line couplers to validate the effectiveness of the ANNs-SaDE approach for broadband design, as illustrated in Fig. 8. The goal is to realize 3-dB wideband couplers with a center frequency adjustable between 2 and 2.5 GHz. The same substrate as in the previous case is used, but with a different thickness of 1.575 mm. The model is trained for 1000 epochs on a dataset of 3000 samples. Testing results from a dataset of 300 samples are summarized in Table III. The SaDE algorithm employs a population size of 100 individuals across 200 generations. Recommended physical parameters are listed in Table IV, and *S*-parameters were validated using Keysight ADS EM solver, as shown in Fig. 9. Table V summarizes the performance of the designed wideband couplers, comparing it with results from ideal cascaded couplers.

TABLE III
TRAINING RESULTS OF THE SURROGATE MODEL FOR WIDEBAND THREE-STAGE COUPLERS

| Electrical Property | MAE in Test Set |
|---|---|
| $|S_{11}|$ (dB) | 0.334 |
| $|S_{21}|$ (dB) | 0.057 |
| $|S_{31}|$ (dB) | 0.030 |
| $|S_{41}|$ (dB) | 0.326 |
| Phase $S_{21}$ (°) | 0.489 |
| Phase $S_{31}$ (°) | 0.443 |

TABLE IV
RECOMMENDED PHYSICAL PARAMETERS OF WIDEBAND THREE-STAGE COUPLERS (ALL IN MM)

| Physical Parameters | W1 | L1 | W2 | L2 | W3 | L3 | W |
|---|---|---|---|---|---|---|---|
| 3 dB @ 2 GHz | 5 | 26.1 | 4.3 | 29.6 | 0.7 | 27 | 5.2 |
| 3 dB @ 2.25 GHz | 5 | 23.4 | 4.2 | 26.6 | 0.7 | 23.9 | 5.1 |
| 3 dB @ 2.5 GHz | 4.9 | 21.1 | 4 | 24.3 | 0.7 | 21.4 | 5 |

TABLE V
COMPARISON RESULTS OF THE DESIGNED COUPLERS WITH THE IDEAL COUPLERS

| Reference | $f_0$ (GHz) | FBW (%) | $k_0$ | Phase Difference (°) |
|---|---|---|---|---|
| Ideal Cascaded Coupler [13] | – | 43 | 0.29 | – |
| Ideal Cascaded Coupler [13] | – | 45 | 0.40 | – |
| Coupler in Fig. 9(a) | 2 | 45.85 | 0.494 | 89.6-91.4 |
| Coupler in Fig. 9(b) | 2.25 | 44.36 | 0.419 | 89.6-91.3 |
| Coupler in Fig. 9(c) | 2.5 | 41.08 | 0.231 | 89.6-91.1 |

Note: $k_0$ represents the coupling imbalance at the center frequency. FBW is calculated for input return loss and isolation levels below -20 dB, with a magnitude imbalance smaller than 0.86 dB.

### IV. CONCLUSION

This paper introduces an ANNs-SaDE-based framework for design automation of branch-line couplers. Using this approach, both folded branch-line couplers and wideband cascaded couplers have been automatically designed, achieving satisfactory performance. Based on the presented results, it can be concluded that the presented approach has the potential to be used for design automation of microwave passive circuits.